# Spatial clustering and common regulatory elements correlate with coordinated gene expression

Jingyu Zhang[1, 2,✦], Hengyu Chen[1,✦], Ruoyan Li[1], David A. Taft[2], Guang Yao[3], Fan Bai[1,*], and Jianhua Xing[2,4,]*

[1]Biomedical Pioneering Innovation Center (BIOPIC), School of Life Sciences, Peking University, Beijing 100871, China

[2]Department of Computational and Systems Biology, School of Medicine, University of Pittsburgh, Pittsburgh, PA, 15260, USA

[3]Department of Molecular and Cellular Biology, University of Arizona, Tucson, AZ, 85721, USA

[4]UPMC-Hillman Cancer Center, University of Pittsburgh, Pittsburgh, PA,15232, USA

✦ These authors contributed equally.

*To whom correspondence should be sent: xing1@pitt.edu, fbai@pku.edu.cn

## ABSTRACT

Many cellular responses to surrounding cues require temporally concerted transcriptional regulation of multiple genes. In prokaryotic cells, a single-input-module motif with one transcription factor regulating multiple target genes can generate coordinated gene expression. In eukaryotic cells, transcriptional activity of a gene is affected by not only transcription factors but also the epigenetic modifications and three-dimensional chromosome structure of the gene. To examine how local gene environment and transcription factor regulation are coupled, we performed a combined analysis of time-course RNA-seq data of TGF-β treated MCF10A cells and related epigenomic and Hi-C data. Using Dynamic Regulatory Events Miner (DREM), we clustered differentially expressed genes based on gene expression profiles and associated transcription factors. Genes in each class have similar temporal gene expression patterns and share common transcription factors. Next, we defined a set of linear and radial distribution functions, as used in statistical physics, to measure the distributions of genes within a class both spatially and linearly along the genomic sequence. Remarkably, genes within the same class despite sometimes being separated by tens of million bases (Mb) along genomic sequence show a significantly higher tendency to be spatially close despite sometimes being separated by tens of Mb along the genomic sequence than those belonging to different classes do. Analyses extended to the process of mouse nervous system development arrived at similar conclusions. Future studies will be able to test whether this spatial organization of chromosomes contributes to concerted gene expression.

## Author Summary

Cellular responses to environmental stimulation are often accompanied by changes in gene expression patterns. Genes are linearly arranged along chromosomal DNA, which folds into a three-dimensional structure. The chromosome structure affects gene expression activities and is regulated by multiple events such as histone modifications and DNA binding of transcription factors. A basic question is how these mechanisms work together to regulate gene expression. In this study, we analyzed temporal gene expression patterns in the context of chromosome structure both in a human cell line under TGF-β treatment and during mouse nervous system development. In both cases, we observed that genes regulated by common transcription factors have an enhanced tendency to be spatially close. Our analysis suggests that spatial co-localization of genes may facilitate the concerted gene expression.

# INTRODUCTION

A cell continuously receives signals from its local environment and accordingly adjusts cellular programs, such as cell proliferation, motility and metabolism [1]. Typically, regulation of a cellular process requires changes in the expression of a group of genes in a temporally coordinated manner [2]. How such coordinated regulation is achieved is a central question that remains poorly addressed.

A mechanism of such regulation is through specific interaction network structures of transcription factors (TFs). TFs bind to certain DNA sites and regulate transcriptional activities of their targeted genes. A TF can regulate multiple target genes to form a so-called single-input-module (SIM, or fan-out) [3]. This SIM network motif appears in a high frequency to coordinate the expression of genes with related functions such as those in bacterial metabolic pathways [4]. Gene regulation in eukaryotic cells is more complex since the three-dimensional structure of DNA has a more profound impact on gene transcription than that in prokaryotic cells. For instance, a nucleosome structure with a high packing level limits gene accessibility [5]. Furthermore, epigenetic modifications can strongly influence gene transcription [6]. It is not fully understood how these different regulation mechanisms collectively control the expression of a group of genes.

To examine how multiple levels of regulation lead to concerted expression of gene groups, we analyzed the temporal gene expression profiles of TGF-β treated human mammary epithelial MCF10A cells in the context of histone modification patterns and chromosome structures derived from Hi-C data. The TGF-β family is crucial for regulating a complex signal transduction network in embryonic and fetal development, and is also involved in multiple physiological and pathological processes such as wound healing and cancer progression [7]. Its signaling event starts from membrane embedded TGF-β receptors, which bind active TGF-β molecules from the extracellular environment [8]. The TGF-β signal is then transmitted into the cell through a signal transduction network and triggers a cascade of cellular responses. The latter is achieved through temporally coordinated expression changes of groups of genes with related functions such as cell proliferation, metabolism, and motility [9]. TGF-β also induces a global reprogramming of cell epigenome [10], which reinforces cellular responses for committed cell phenotype transition. We also analyzed temporal gene expression together with histone modifications and chromosome structures during mouse neural differentiation, another well-defined model for studying cell phenotype transition [11, 12]. Specifically, we analyzed a recently published dataset that combined Hi-C, RNA-seq, and ChIP-seq studies on the differentiation process from mouse embryonic stem cells (ESCs) to neural progenitor cells (NPCs) then to cortical neurons (CNs) [13]. In both the TGF-β response and neural differentiation systems, our analyses reveal that genes co-regulated by a common TF(s) have the tendency to be spatially close, even if they are distant along the linear genome sequence.

## MATERIALS AND METHODS

### Cell culture

MCF10A cells were purchased from the American Type Culture Collection (ATCC) and were cultured in the DMEM/F12 (1:1) medium (Gibco) with 5% horse serum (Gibco), 100 μg/ml of human epidermal growth factor (PeproTech), 10 mg/ml of insulin (Sigma), 10 mg/ml of hydrocortisone (Sigma), 0.5 mg/ml of cholera toxin (Sigma), and 1x penicillin-streptomycin (Gibco). Cells were cultured at 37 °C with 5% $CO_2$ with a medium change every the other day. We induced the cells with 4 ng/ml human recombinant TGF-β1 (Cell signaling).

### RNA extraction and library preparation

Total RNA was isolated from the cell pellets with an RNA extraction kit (Qiagen, Cat No. 74104). All RNA extracts were confirmed with high quality (RQN score = 10.0) using the Fragment Analyzer$^{TM}$ platform (Advanced Analytical Technologies, Inc). Libraries were prepared using the NEBNext Ultra RNA Library Prep Kit for Illumina (NEB, Cat No. E7530L) according to the manufacturer's instructions. Briefly, mRNA was first isolated from total RNA with oligo d(T)25 beads (all volumes were halved except for washing steps, NEB, Cat No. E7490S). Next, purified mRNA was denatured and melted into small fragments, and subjected to random priming and extension for reverse transcription. After that, double-stranded cDNA was end-repaired, dA-tailed, adaptor ligated, and amplified with 12 PCR cycles. Constructed libraries were subjected to purification and quality control; the final quality-ensured libraries were pooled and sequenced on an Illumina HiSeq 4000 instrument for 150 bp paired-end sequencing.

**RNA-seq data processing**

Paired-end cleaned reads were aligned to the human reference genome hg19 (UCSC) using TopHat (v 2.1.1) with default parameters. The BAM files of mapped reads were used to annotate transcripts and calculate the FPKM values using the Cufflinks, Cuffquant, Cuffnorm suite [14]. Differentially expressed (DE) genes were identified between any two time points with the criteria: fold change >2 or < 0.5 and FDR < 0.05. The FPKM values of genes from the RNA-seq dataset were further cleaned up using custom R scripts. Hierarchical clustering of genes was performed using an R package (pheatmap). Gene expression and TF regulation based Hidden Markov Model (HMM) clustering was performed with the DREM2 software [15]. RNA-seq results of ESC, NPC and neuron cells were downloaded from the GEO database under the accession number GSE96107.

**Chromosome structure analyses**

Hi-C data were downloaded from the GEO database (MCF10A, GEO:GSE66733; mouse nervous system GEO:GSE96107). Chromosome structures were constructed using an R package (igraph). Clustering of bins was obtained with the fast-greedy algorithm [16]. Physical distances between bins were estimated with a Matlab code provided by Lesne et al. [17]. This code uses a Shrec3D algorithm, which first relates the Hi-C contact frequency between every two genomic sites with a spatial distance, then approximates the actual distance between the two sites by their shortest-path distance on a contact graph. This algorithm alleviates uncertainty of reconstructing the spatial distance between two distal sites only by their own contact frequency.

**Distribution function calculation**

**Linear distribution function:** For a tagged HMM class α gene, we divided the flanking sequences into bins with a size of Δl base pairs, and the *i*-th pair of bins [(−i − 1)Δl, −i Δl] and [i Δl, (i + 1)Δl], i = 0, 1, etc. (Fig S1A). In the *i*-th pair of bins, there are $n_{i\alpha}$ genes belonging to the same HMM class as the tagged gene. For the *0*-th pair of bins the counting of the genes should exclude the tagged gene. The linear correlation was calculated as $\sigma_\alpha^L(i) = \frac{\langle n_{i\alpha}+n_{-i\alpha}\rangle_\alpha}{N_\alpha-1}$, where $i$ = 0, 1, 2, etc. $N_\alpha$ was defined as the total number of genes belonging to class α, and the average $\langle\bullet\rangle_\alpha$ was performed over every HMM class α gene as the tagged gene.

As a control,

$$\sigma_\alpha^{LA}(i) = \frac{\langle n_i+n_{-i}\rangle_\alpha}{N-1},$$

where $n_i$ was the total number of genes in the $i$-th bin and $N$ the total number of human genes,

and

$$\sigma_\alpha^{LD}(i) = \frac{\langle n_{iD}+n_{-iD}\rangle_\alpha}{N_D-1},$$

where $n_{iD}$ is the total number of DE genes in the $i$-th bin, and $N_D$ is the total number of DE genes.

**Spatial distribution function**: The idea of a radial distribution function from statistical mechanics was implemented (Fig S1B) [18]. Each chromosome was divided into sequence bins with a size of 250 kb. A tagged gene from HMM class α resides in a bin that we referred to as the tagged bin. The spatial distance in the 3D physical space between the tagged bin and another bin containing a specific HMM class β gene was analyzed using the Shrec3D algorithm [17] to convert the contact frequency between two bins from Hi-C data to a spatial distance. The sphere centered at the tagged bin was divided into shells with a thickness Δr. In our analysis, Δr ≈ 60 nm based on the estimated conversion in [17].

Next, an average spatial correlation function between a class-α-gene-containing bin at the origin and class-β-gene-containing bins within the $i$-th shell was defined as,

$$\sigma_{\alpha\beta}^R(i) = \langle\frac{n_{\beta i}}{\frac{N_\beta}{V}\frac{4}{3}\pi[((i+1)\Delta r)^3-(i\Delta r)^3]}\rangle_\alpha, i = 0, 1, \text{etc.},$$

where $n_{\beta i}$ is the number of HMM class β genes within a spherical shell $(i\Delta r, (i+1)\Delta r)$, and this number excludes the tagged class α gene within the $0$-th shell in the case $\beta = \alpha$;

$N_\beta$ is the total number of class $\beta$ genes and this number excludes the tagged class $\alpha$ gene in the case $\beta = \alpha$; $V$ is the volume of the nucleus and the unit was chose so that $V = 1$, and the average over $\alpha$ is again performed over all genes belonging to class $\alpha$ as the tagged gene.

Similarly, the controls were defined as,

$$\sigma_{\alpha A}^R(i) = \langle \frac{n_i}{\frac{(N-1)}{V} \frac{4}{3}\pi[((i+1)\Delta r)^3 - (i\Delta r)^3]} \rangle_\alpha,$$

$$\sigma_{\alpha D}^R(i) = \langle \frac{n_{iD}}{\frac{(N_D - 1)}{V} \frac{4}{3}\pi\left[((i+1)\Delta r)^3 - (i\Delta r)^3\right]} \rangle_\alpha,$$

where $n_i$ is the number of all genes within the $i$-th shell; $N$ is the total number of human genes; $n_{iD}$ is the number of DE genes within the $i$-th shell; and $N_D$ is the total number of DE genes. Again, the tagged gene was excluded when counting $n_0$ and $n_{d0}$.

## RESULTS

### Changes in gene expression reflect cell phenotype transition in response to TGF-β

We used MCF10A cells, a non-tumorigenic human mammary epithelial cell line, as a major *in vitro* model to study in this work. This cell line has been widely used to study the TGF-β induced epithelial-to-mesenchymal transition (EMT) [1, 19] (Fig 1A). Cells were treated with 4 ng/ml TGF-β for 12 hours, 2, 3, 5, 8, 12, and 21 days (Fig 1B). Untreated MCF10A cells showed typical epithelial morphology with tight cell-to-cell adherence. With TGF-β treatment, we observed progressive morphological changes indicating the transformation from epithelial to mesenchymal phenotype. From day 2 to day 5, cells started to show loosened intercellular adherence. After day 5, some cells appeared with expanded cell size and extended long cell axis. With further TGF-β treatment, more cells acquired a spindle-like shape. On day 21, only a small fraction of cells still maintained epithelial morphology and most cells had undergone EMT.

Next, we performed RNA-seq studies to uncover changes of gene expression accompanying EMT. At each time point, we harvested cell samples and extracted RNA. The RNA-seq results revealed that about 33% of human genes were differentially expressed upon TGF-β treatment. Principal component analysis (PCA) over these ~ 7000

DE genes showed an expected larger separation between gene expression profiles of samples from different time points than those of replicate samples from the same time point (Fig 1C). The global transcriptome change over time reflected in the PCA space was consistent with the gradual morphological change of cells over time and the previous report that TGF-β-induced EMT proceeded through intermediate states [19].

## Gene classes sharing similar expression patterns and upstream regulators exhibit similar functional characteristics

To further examine the temporal patterns and functions of the DE genes, we performed hierarchical clustering (HC) analysis. The analysis divided the DE genes into seven HC classes based on similar expression patterns in each (Fig 2A) [20]. Among the seven HC classes, class I with ~1,700 genes exhibit a monotonically decreasing pattern, and class II of ~2,000 genes exhibit a monotonically increasing pattern. Another two classes III and IV show transient up and transient down dynamics, respectively. The remaining three classes V-VII display wavy dynamic patterns to varying degrees. Gene ontological (GO) analysis (Fig S2) revealed that genes in each class are typically involved in multiple cellular processes. For example, genes in the decreasing class (class I) are related to RNA polymerase I activity and snoRNA binding. These two functions are related to the RNA metabolic process, including ribosomal RNA production, modification, and binding to regulatory factors. The observation that these genes are down-regulated is consistent with previous reports that under TGF-β treatment cells are under growth arrest until they finish EMT [21].

Histone modifications can also affect gene expression [22]. To investigate the relationship between histone modification and gene expression, we integrated genome-wide H3K4me3 and H3K4ac profiles obtained by Messier et al. [23] with our RNA-seq data. Both H3K4me3 and H3K4ac are histone modification marks that are associated with active or poised genes [24]. We used H3K4me3 and H3K4ac profiles of all human genes as a control, and examined the marks in each HC class. The results in Fig 2B show that all HC classes have elevated H3K4me3 and H3K4ac compared to the control, and there is no apparent difference between different classes. Each HC class also has a broad bimodal distribution. That is, genes within an HC class do not share common histone modification patterns. Given that histone modification patterns correlate with local chromosome structures [25], these results suggest that genes from the same HC class have heterogeneous local chromosome environments.

Next, we adopted a different clustering scheme, the Dynamic Regulatory Events Miner (DREM), which clusters genes by combining gene expression time series with additional pre-established transcriptional networks [26]. Figure 3A shows the clustering results analyzed with DREM2 based on a Hidden Markov Model (HMM) [15]. At each conjunction node, genes are assigned to different branches based on their expression trend and upstream regulators (transcription factors on this node). Genes from an upstream branch can become key regulators at subsequent nodes [15, 26]. It reveals a hierarchy of gene regulation during the process of TGF-β-induced phenotype change. With DREM2 the DE genes were clustered prominently into 46 branches with 19 nodes at the conjunction sites and 25 end classes. For clarity, we call the latter HMM classes to distinguish from the HC classes that are based on expression only.

Compared to the HC classes, HMM classes showed finer dynamic patterns and GO enrichment information (Table S1). For example, genes in the first seven HMM classes all had increased expression, but differed in their detailed temporal profiles. Genes in class C1 increased their expression to high levels already on day 2. Genes related to metalloendopeptidase activity were enriched in this class by over 17 fold with respect to the reference genes. Four of the matrix metalloproteinases (*mmp*s), *mmp2/7/11/13*, are also in this class. These four MMPs are known to degrade components of extracellular matrix proteins such as gelatin, fibronectin, and laminin, and mediate biological activities including migration, mammary epithelial cell apoptosis, and EMT [27]. Heparin binding genes were another type of highly enriched genes. These genes, such as *periostin* (*postn*), *fibronectin* (*fn1*), are also known to be related to matrix or cell membrane formation and thus affect cell migration and adhesion [28]. Another class of early activation genes, class C2, was also enriched with genes related to cell matrix and membrane structure. Among them five of the *pcdh* family members, including *pcdh7/a4/b9/b10/b13*, are integral membrane proteins that are involved in cell-cell recognition and adhesion [29]. In general, genes within each HMM class had narrower distributions and thus higher similarity of histone modification patterns (Fig 3B) than those of the HC classes do (Fig 2B).

Therefore, genes clustered through the DREM2 analysis based on common TFs and similar dynamic profiles tend to have closely related functions.

**Genes sharing common regulators have an enhanced tendency to be spatially close**

As mentioned above, local chromosomal DNA environment affects gene transcriptional activity. We wondered whether genes sharing similar expression patterns and common regulatory factors, as in an HMM cluster identified by the DREM2 analysis, are also spatially close and share similar local DNA environment. To test this hypothesis, we first examined gene arrangement along the linear genome sequences. We divided the whole human genome into bins with a resolution of 1 Mb, a typical size of a topologically associated domain (TAD). Then we matched all genes to the relevant bins based on their genomic positions. Statistical analysis of all the genes spreading along the chromosomes showed that genes are not evenly distributed along the DNA sequences (Fig 4A). Most bins have less than ten genes, and globally one third of the bins are gene-free. By contrast, ~3% of the bins (a total of less than 100 bins) contained 17% of the overall human genes. This uneven distribution was slightly more profound for the DE genes under TGF-β treatment: DE genes resided in less than half of the bins and 17% of DE genes were enriched in only 2.5% of the bins.

To further examine the gene distribution along chromosomes, we defined an averaged linear distribution function $\sigma^L$ (see Materials and Methods for details). It measured how the chromosomal density of a group of genes of interest changes with respect to the transcription starting site (TSS) of a given gene. For a given gene $x$ belonging to an HMM class α as a tagged gene, we divided the DNA sequences along both sides flanking the TSS of $x$ into bins with a size of 125 kb (Fig S1A, r = 125 kb), and counted the fraction of HMM class α genes in each bin. We then repeated this process by choosing every gene in the HMM class α as the tagged gene, and calculated the average density of HMM class α genes ($\sigma^L_\alpha(i)$) in the $i$-th bin with respect to the tagged gene. For statistical comparison, we also calculated a similar $\sigma^{LA}_\alpha(i)$ for all human genes and $\sigma^{LD}_\alpha(i)$ for all DE genes with respect to the tagged HMM class α genes as controls. If there were no class-specific gene clustering along the genomic sequence, one would expect that $\sigma^L_\alpha(i) = \frac{\langle n_{i\alpha}+n_{-i\alpha}\rangle_\alpha}{N_\alpha-1} = \frac{\langle n_i+n_{-i}\rangle_\alpha}{N_\alpha-1}\frac{N_\alpha-1}{N} = \sigma^{LA}_\alpha(i) = \sigma^{LD}_\alpha(i)$ within statistical errors (see

Materials and Methods for explanation of terms). Instead, the $\sigma^L$ values of more than half of HMM classes were not significantly higher than those of DE gene and all human gene controls. The upper left panel of Fig 4B shows HMM class C23 as an example. Only five HMM classes showed statistically significant increases of $\sigma^L$ values over controls (although the increases are small) within the first pair of bins ($\leq$ 125 kb), indicating relative accumulations of genes from the same HMM class; one of them (HMM class C24) is shown in Fig 4B upper right panel.

Next, we investigated the spatial arrangement of the DE genes using a set of available Hi-C data from MCF10A cells [30]. Following an approach used in statistical mechanics [18], we defined a set of radial distribution functions ($\sigma^R_{\alpha\beta}(i)$) that measured the average radial density (rd) of HMM class β genes and residing inside the *i*-th evenly divided spherically shell relative to a tagged class α gene, and averaged the rd values over all class α genes (Fig S1B). For comparison we also defined two controls $\sigma^R_{\alpha A}(i)$) and $\sigma^R_{\alpha D}(i)$), where the class β genes were replaced by all human genes and all DE genes, respectively. If there were no HMM class-specific gene spatial clustering, one would expect that within statistical errors, $\sigma^R_{\alpha\alpha}(i) = \sigma^R_{\alpha\beta}(i) = \sigma^R_{\alpha A}(i)) = \sigma^R_{\alpha D}(i)$. According to this metric, however, genes in the classes C23 and C24, as discussed above, exhibited substantial spatial clustering. Genes in class C24 tended to be spatially close (Fig 4B bottom right), likely due to their arrangement in a linear sequence. Notably, genes in class C23 also showed significantly enhanced spatial co-localization. With respect to a tagged C23 gene, the rd values of C23 genes within the first shell was more than doubled than that of all genes, which means that even some C23 genes that are not close along the linear sequence come close spatially. To visualize such spatial clustering of genes from an HMM class, we generated a two-dimensional plot of 1-Mb bins on chromosome 1 based on bin-bin contact frequencies obtained from the Hi-C data (Fig 4C). The red boxes show spatial aggregation of genes on chromosome 1 that belong to the two classes, respectively. Further analysis revealed significant gene spatial clustering in the first shell for all HMM classes compared to that of the controls (Fig 4D), and showed that spatial clustering mainly takes place within each HMM class (Fig S3A). That is, genes sharing a common upstream regulator have an enhanced tendency to be spatially close.

We also examined how the genes within the first shell of a tagged gene are distributed along the chromosome sequence (Fig S3B). While a large contribution to the average radial gene density ($\sigma_{\alpha\alpha}^R(0)$) came from genes that were already close along the chromosome sequence, some gene elements as far as ~ 50 Mb apart resided spatially close.

**Genes of similar functions tend to cluster spatially during embryonic development**

Next, we asked whether the observed spatial clustering of genes with related function is beyond the TGF-β induction of MCF10A cells. To this end, we performed similar DREM2 and linear/spatial gene density analyses on the differentiation of mouse ESCs into NPCs then CNs (Fig 5A), for which both RNA-seq and Hi-C data for the three developmental stages were reported by Bonev et al. [13]. A DREM2/HMM analysis clustered ~ 20,000 mouse genes into seven classes based on both their expression patterns during neuron cell differentiation and TF regulation (Fig 5B). For both ESC and CN cells, radial distributions (Fig 5C, 5D, and Fig. S4A) show that genes within the same HMM class have a slightly enhanced tendency to cluster spatially in the first shell compared to the control groups. We observed a similar tendency for NPC cells but to a less extent.

Compared to the MCF10A cell data, the ESC-CN system showed less enhanced spatial clustering within individual HMM classes relative to that of the control. We reasoned that DREM2 clustering was more coarse-grained in the ESC-CN system due to the limited number of time points in the available RNA-seq datasets. Each HMM class is thus likely composed of multiple sub-classes regulated by different TFs. The expected effect of spatial clustering within each sub-class ($\sigma_{\mu\mu}^R(0)/\sigma_{\mu A}^R(0) > 1$) is then reduced by their spatial relation to other sub-classes ($\sigma_{\mu\nu}^R(0)/\sigma_{\mu A}^R(0) \approx 1$), where μ and ν refer to two sub-classes within one HMM class. Apparently, the ratio reaches an asymptotic value of one if there is only one HMM class. This reduction due to unresolved class mixtures was less severe for MCF10A cells, for which the DREM2 clustering was finer. To support this hypothesis, we reanalyzed the MCF10A RNA-seq data assuming that one can only identify nine HMM classes branched on day 2 (Fig S4B), and eight of them are mixtures of the finer classes obtained from analyzing RNA-seq data at all time points (as shown in Fig 3A). As expected, Fig S4C shows that the extent of spatial clustering of genes within each class is reduced as compared to those shown in Fig 4D.

**Discussion**

Recent studies on chromosome conformations have revealed the existence of structural units such as promoter-enhancer hubs, topologically associated domains (TADs), and meta-TADs and demonstrated that these structural units play important roles in gene regulation [31-34]. Several studies focusing on specific genomic regions have shown correlation between gene expression and local chromosome structures [35, 36]. In this work, we provide a genome-wide perspective on the relationship between chromosome structure and gene regulation by integrating the RNA-seq and Hi-C data. We first only used the expression data and grouped genes that share similar temporal expression patterns and are co-regulated by common TFs together. We found that genes within each group display a significantly enhanced tendency to be clustered spatially in the three-dimensional chromosome structure, regardless whether these genes are close (< 1 Mb) along the genome sequence or separated by as far as tens of Mb. This observation further suggests that the three-dimensional chromosome structure is part of a multi-layer gene regulation program.

Our analysis reveals two related mechanisms that achieve spatial clustering of genes subject to common regulators. Some genes are located close in chromosome sequence and consequently spatially close. By contrast, some genes that are far apart along chromosome sequence can also become adjacent spatially by forming three-dimensional structures. TFs may actively orchestrate such chromosome structure organization [37, 38]. Alternatively, other DNA binding factors such as long non-coding RNAs and transcription initiation complexes can drag associated chromosomal regions together to form enhancer-promoter hub structures. These hub structures may facilitate TF binding and related cooperative regulation such as phase separated molecular assemblies [39].

Functionally, spatial co-localization may contribute to temporally coordinated regulation of a group of genes in eukaryotic cells. This co-localization can be viewed as a further refinement of the SIM network motif first noticed in prokaryotic cells. Spatial co-localization may facilitate simultaneous regulation of local chromosomal environment of these genes, such as DNA methylation and histone modification, and chromosome compaction, all of which affect gene expression activities. Indeed, a recent study on Drosophila embryos shows that a group of genes separated by genomic distance but pulled together by an enhancer element exhibit similar expression fluctuation patterns [40].

Our analysis of the MCF10A data, however, has a number of limitations. While we performed RNA-seq analysis of MCF10A cells at a number of time points during TGF-β treatment, the lack of simultaneous time-course Hi-C and epigenomic data prevented us from analyzing how spatial clustering may change dynamically upon the change of gene expression status. In addition, having the RNA-seq, Hi-C and epigenomic datasets obtained from different labs also raises a concern of potential cell line drifting during culture. It is desirable to have an integrated set of parallel RNA-seq, epigenomic and Hi-C measurements from the same batch of cells, similar to how the ESC differentiation was studied by Bonev et al. [13] but at more time points. Together with the gene regulatory network analysis, such datasets would permit finer clustering and identifying gene groups that each contains multiple spatially clustered, co-regulated and functionally related genes, and examining to what extent these units are either cell type specific or conserved among different cell types.

In summary, based on an integrated analysis of transcriptome, epigenome, and chromosome 3D structural information we propose a mechanism for concerted regulation of gene groups that can be further evaluated with more systematically measured datasets. That is, concerted gene regulation can be achieved through a common trans regulator(s) and the spatial co-localization of target genes. This observation further suggests that genes may be spatially organized into functional units, consistent with the hierarchical patterns and long-range interactions revealed by chromosome structure studies [36, 41]. The relationship between gene expression and chromosome structure can be better understood by grouping genes into finer HMM classes based on their expression patterns and regulatory elements.

## ACKNOWLEDGEMENT

This work was supported by the National Science Foundation [DMS-1462049 to JX,

**Captions**

**Figure 1 MCF10A cell responses to TGF-β treatment.**

**(A)** Schematic diagram of phenotypic transition from epithelial cells to mesenchymal cells in response to TGF-β treatment. **(B)** MCF10A cells undergo morphological changes in responses to 4 ng/ml TGF-β. **(C)** PCA clustering of TGF-β treated MCF10A cells reveals distinct gene expression patterns over time.

**Figure 2 TGF-β induced gene expression changes show distinct temporal patterns.**

**(A)** Hierarchical clustering of genes based on temporal gene expression patterns only. **(B)** Violin plots of distributions of indicated histone modification levels sampled through genes belonging to individual hierarchical clusters. Numbers 1-7 on the x-axis follow the order of gene clusters in panel A. The control group 'A' was sampled through all genes.

**Figure 3 Genes clustered based on both expression patterns and key transcription factors show a correlation between patterns of expression and histone modification.**

**(A)** Dynamic regulatory map obtained through the DREM2 analysis. **(B)** Distribution of indicated histone modification levels sampled through genes belonging to individual HMM classes. Group 'A' represents the control group that includes all genes.

**Figure 4 Genes with similar expression patterns and controlled by the same up-stream regulators show an enhanced tendency to co-localize spatially in the 3D chromosome structure.**

**(A)** Heat map shows the numbers of 1-Mb bins containing a given number of genes and TGF-β responding genes. The orange line highlights the bins in which all genes responded to TGF-β treatment. **(B)** Linear and radial distribution functions of TGF-β-responding genes within two representative HMM classes. We calculated the distribution of genes by sampling three types of gene groups: all available genes (All genes, $\sigma_\alpha^{LA}$ and $\sigma_{\alpha A}^R$), genes within an indicated HMM group (HMM genes, $\sigma_\alpha^L$ and $\sigma_{\alpha\alpha}^R$), and genes that showed differential expression during TGF-β treatment (DE genes, $\sigma_\alpha^{LD}$ and $\sigma_{\alpha D}^R$). For spatial distance, with the shell width Δr approx 60 nm. **(C)** Pseudo-spatial arrangement of genes belonging to two representative HMM classes, respectively. Each circle indicates a 1-Mb bin. The gray level in a circle scales to the number of genes in the bin that belongs to the indicated HMM class. The two-dimensional spatial arrangement of bins within one chromosome was calculated by a fast-greedy algorithm based on the contact frequencies between each pair of bins from Hi-C data. The line width between two circles is proportional to the contact frequencies between the two corresponding bins. Genes within each red box are within ~ 300 nm in space. **(D)** Relative gene densities of all HMM classes within the first shell of the radial distributions normalized by the average density of all genes around the targeted genes in the first shell (i.e., $\sigma_{\alpha\alpha}^R(0)/\sigma_{\alpha A}^R(0)$ and $\sigma_{\alpha D}^R(0)/\sigma_{\alpha A}^R(0)$).

**Figure 5 Mouse ESC-CN system shows a similar enhanced tendency of physical proximity for co-regulated genes with similar expression patterns.**
**(A)** Schematic diagram of the development from ESC to CN cells. **(B)** Dynamic regulatory map obtained through the DREM2 analysis. (**C**) Heat map of intra- (diagonal) and inter- (off-diagonal) HMM class gene densities within the first shell of radial distribution relative to the corresponding densities of all genes (as control) in ESC cells or CN cells (i.e., $\sigma_{\alpha\beta}^R(0)/\sigma_{\alpha A}^R(0)$). (**D**) Gene densities of all HMM classes within the first shell of radial distribution relative to the corresponding densities of all genes (i.e., $\sigma_{\alpha\alpha}^{R}(0)/\sigma_{\alpha A}^{R}(0)$)) in ESC cells or CN cells.

Fig. 1

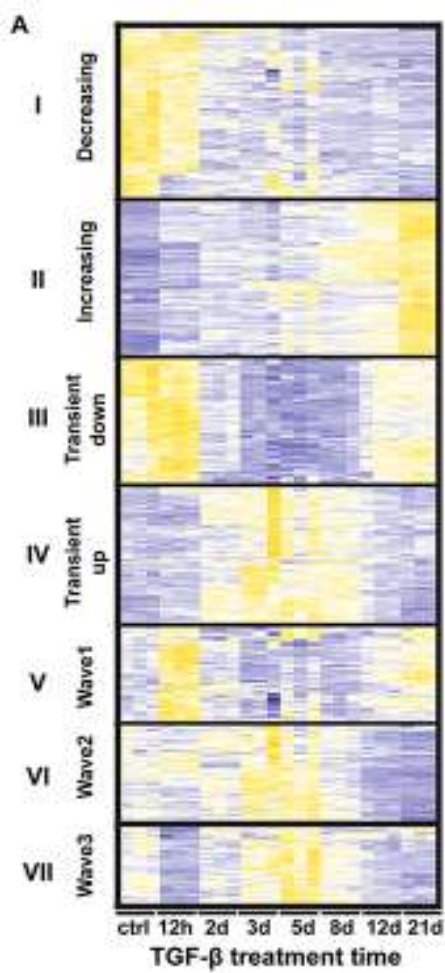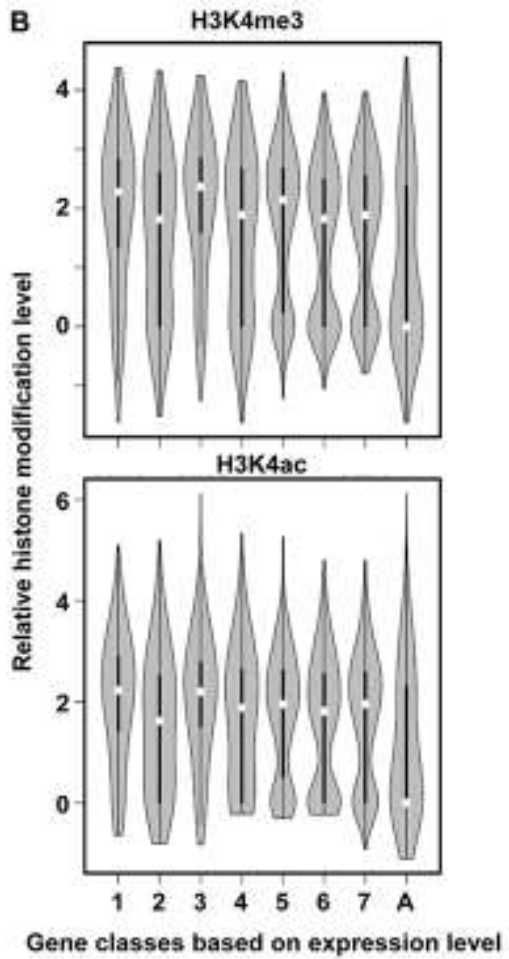

Fig. 2

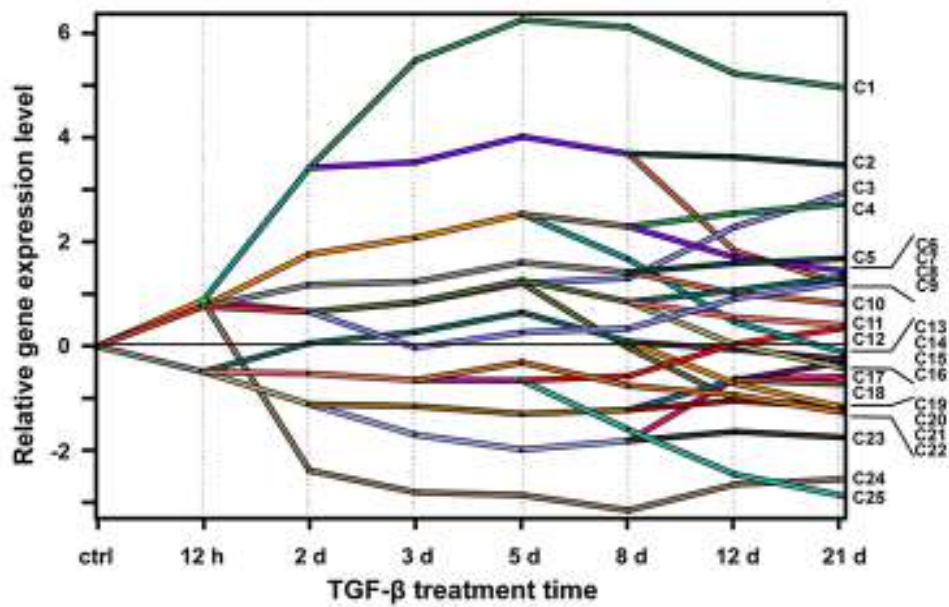
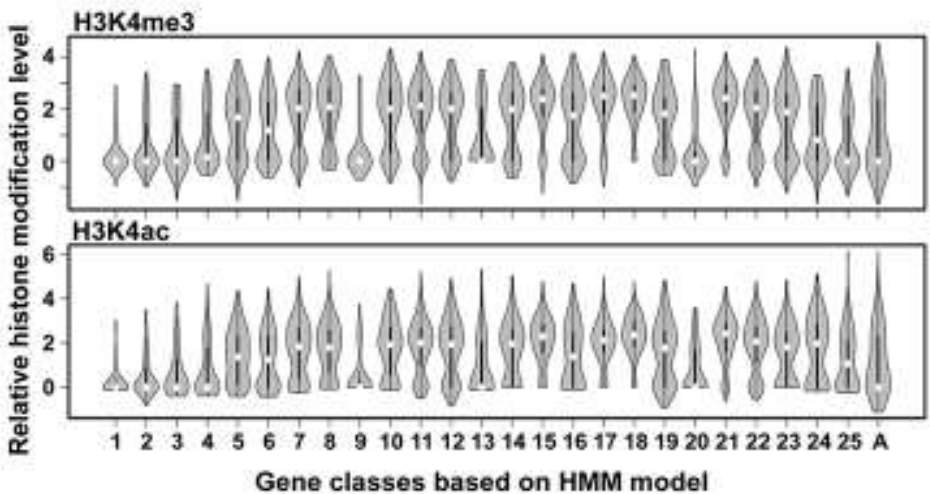

Fig. 3

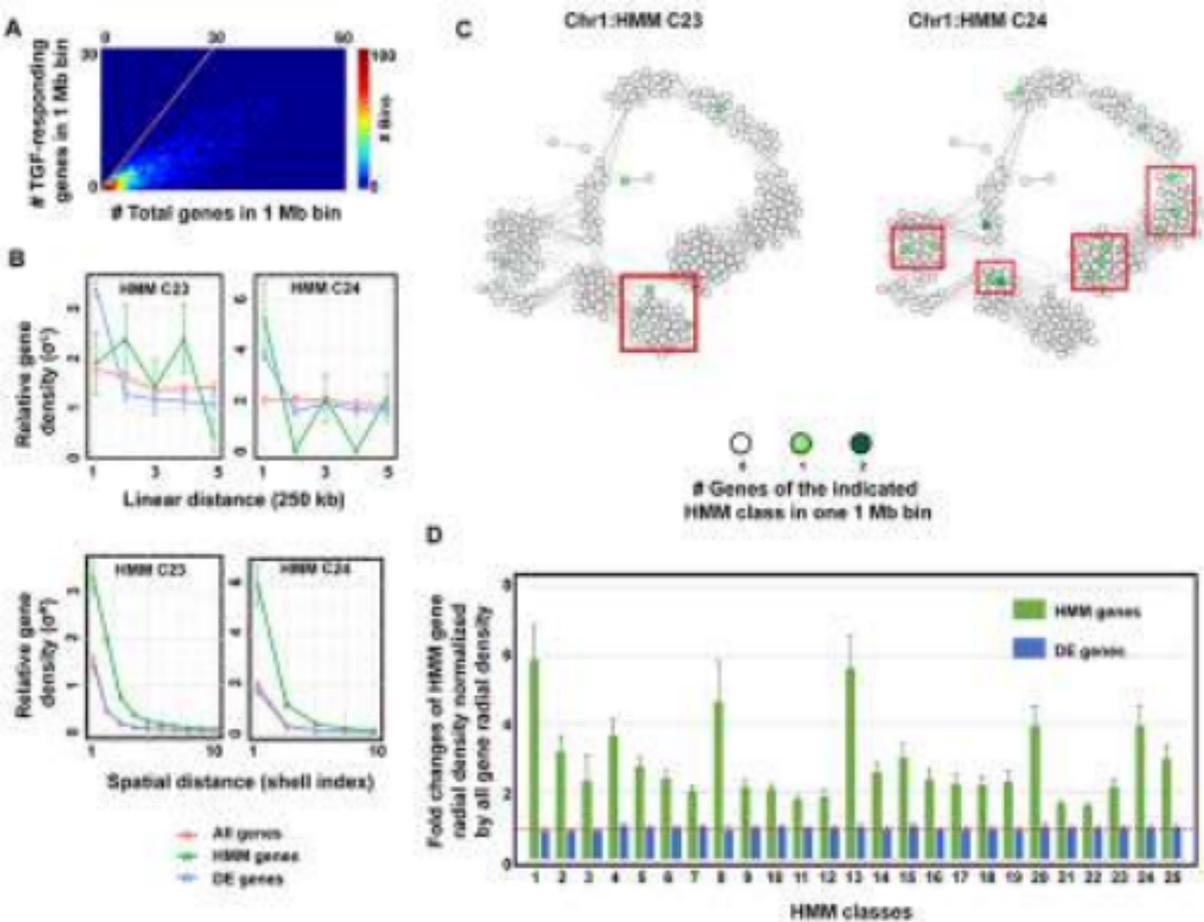

Fig. 4

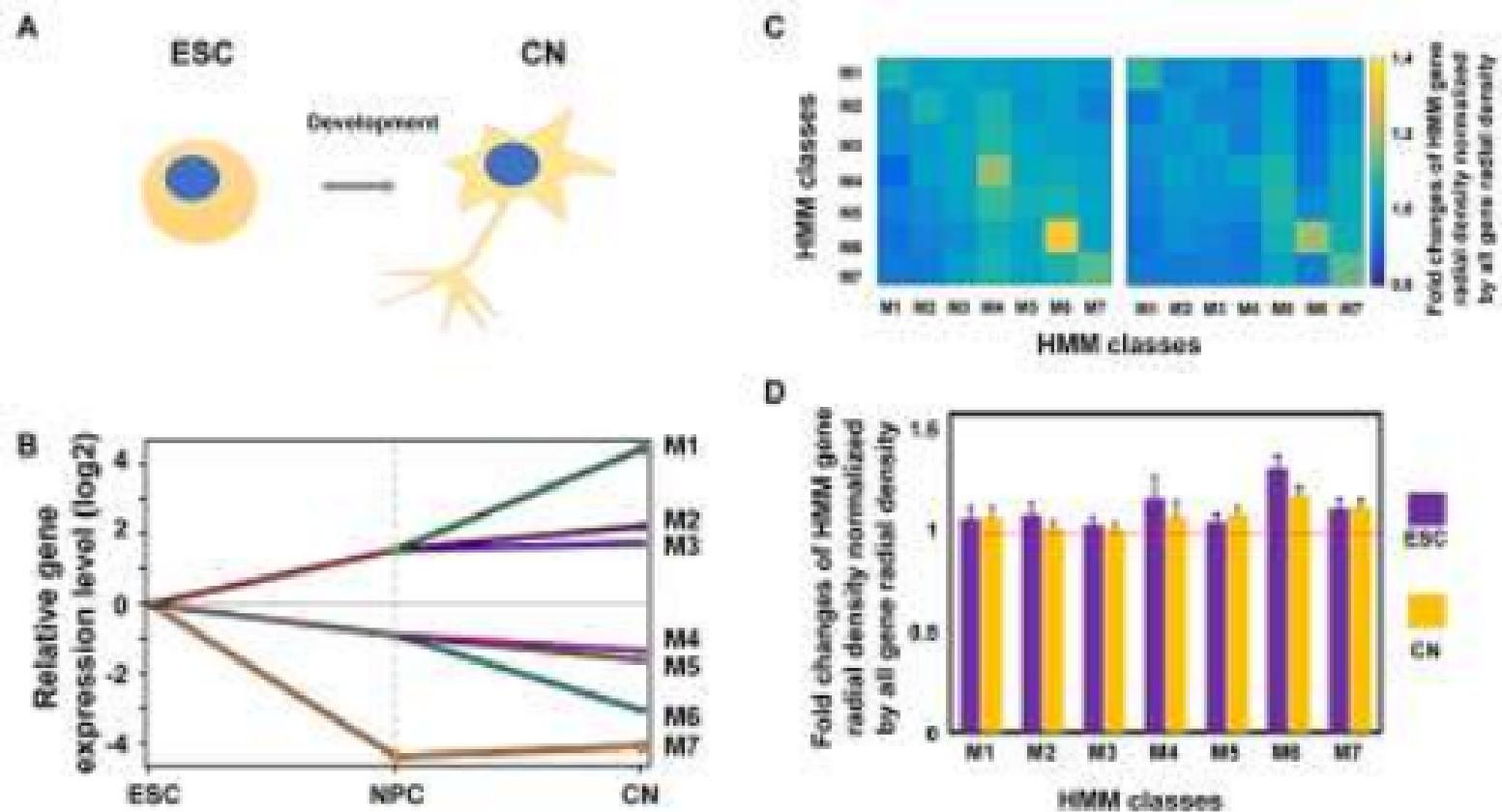

Fig. 5

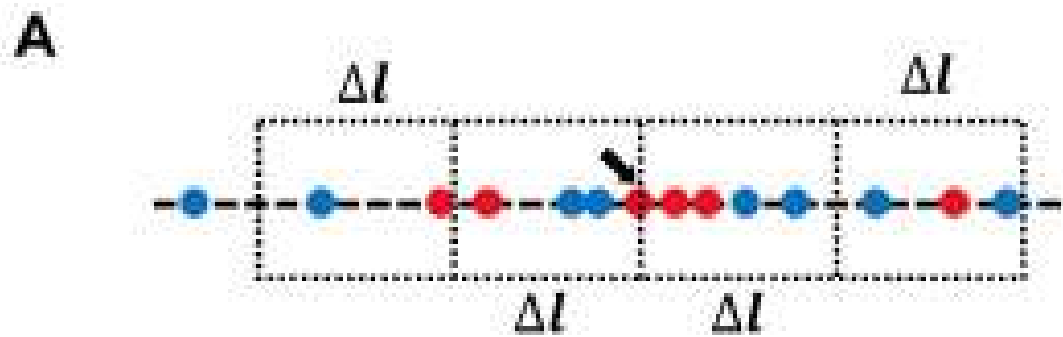
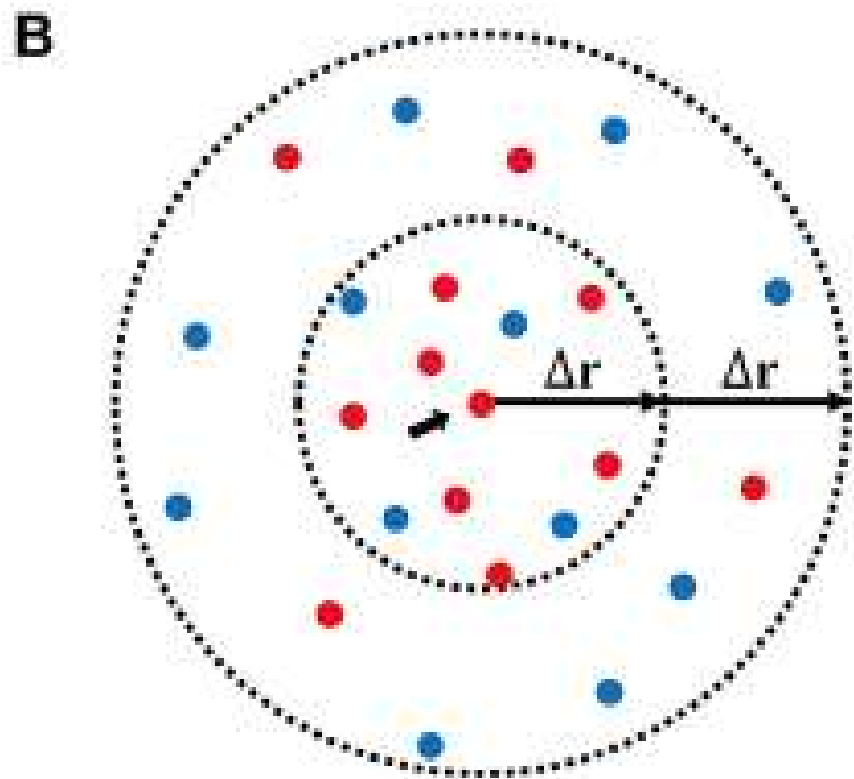

**Fig. S1**

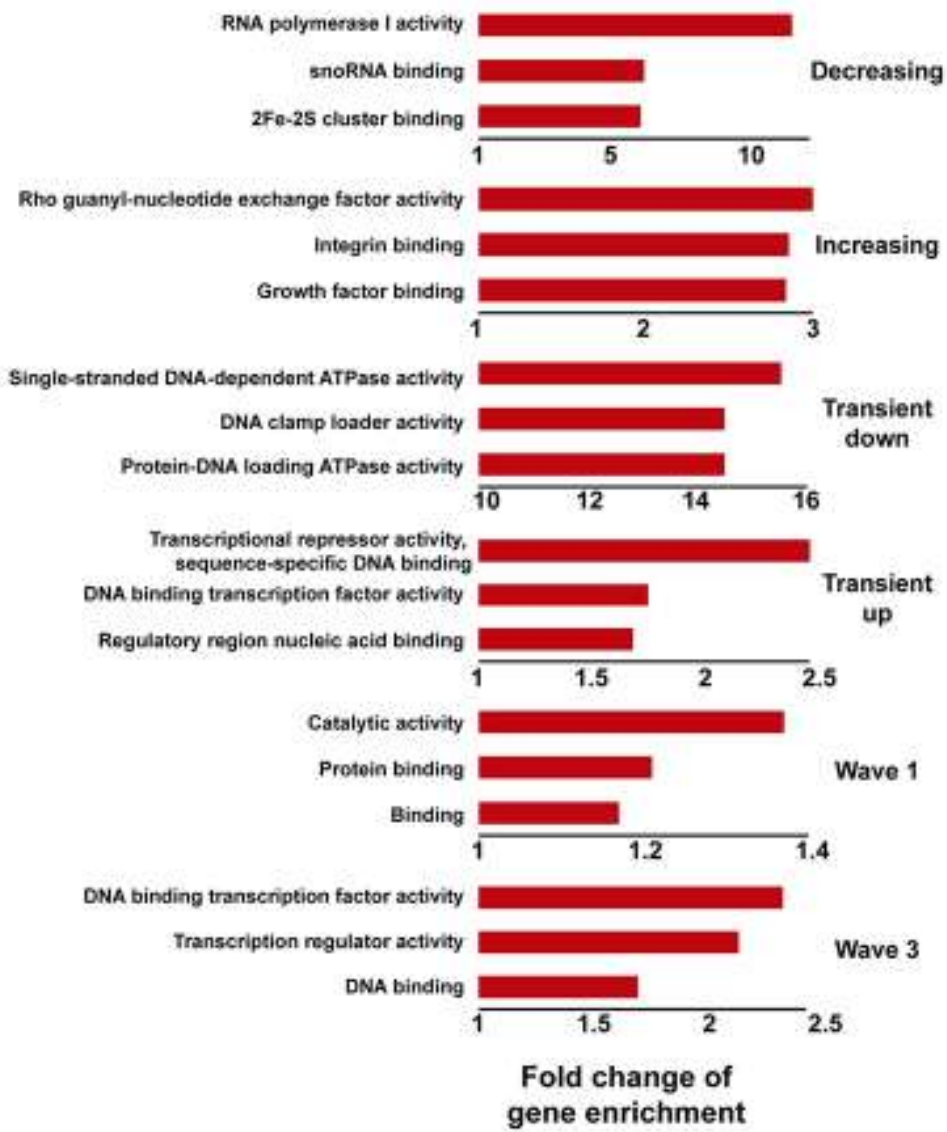

Fig. S2

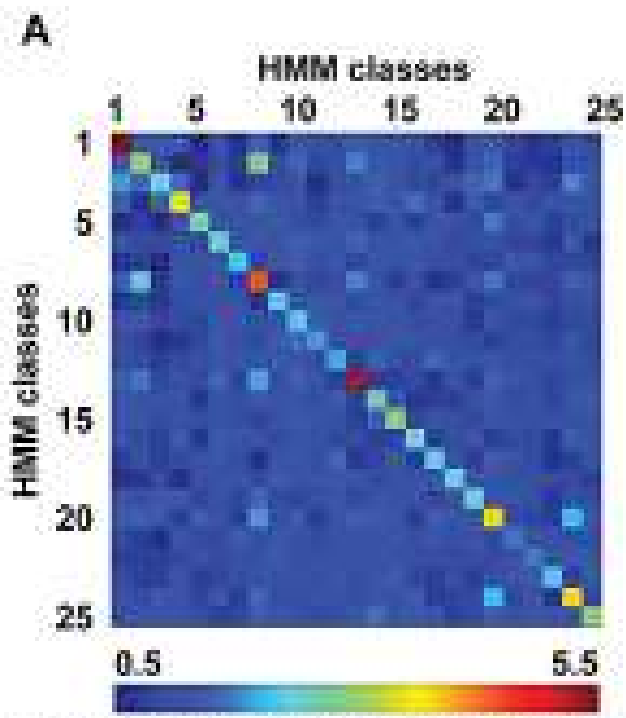
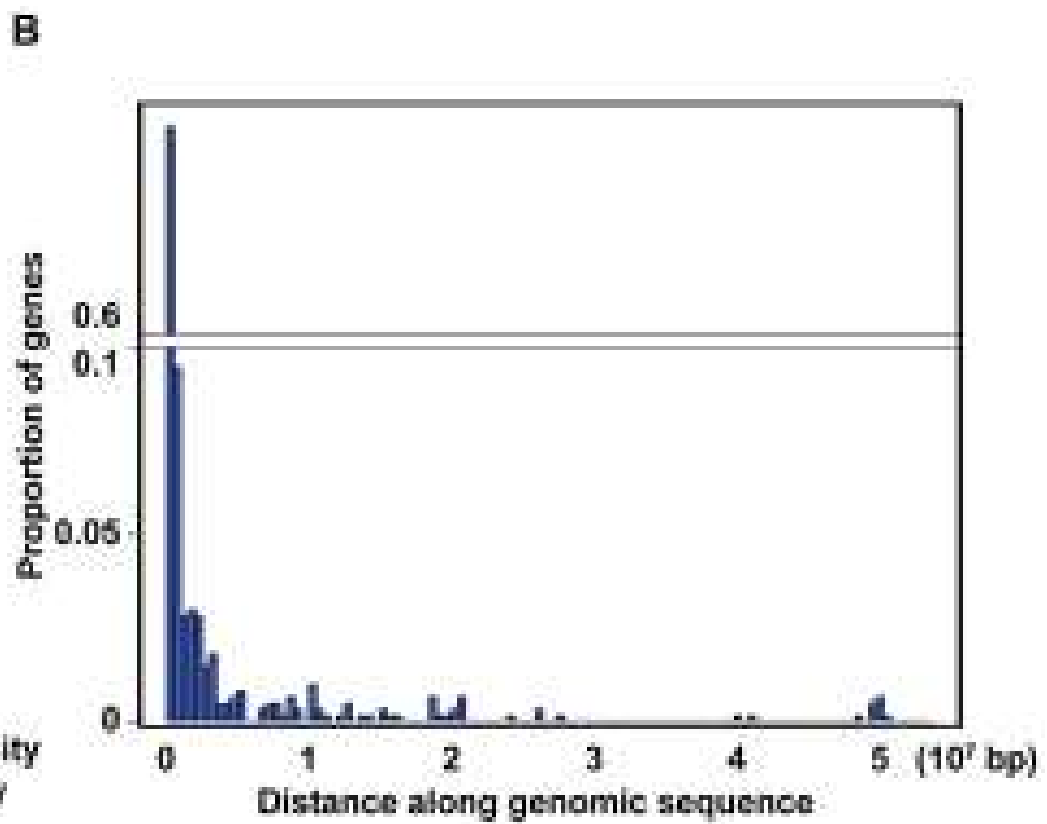

Fig. S3

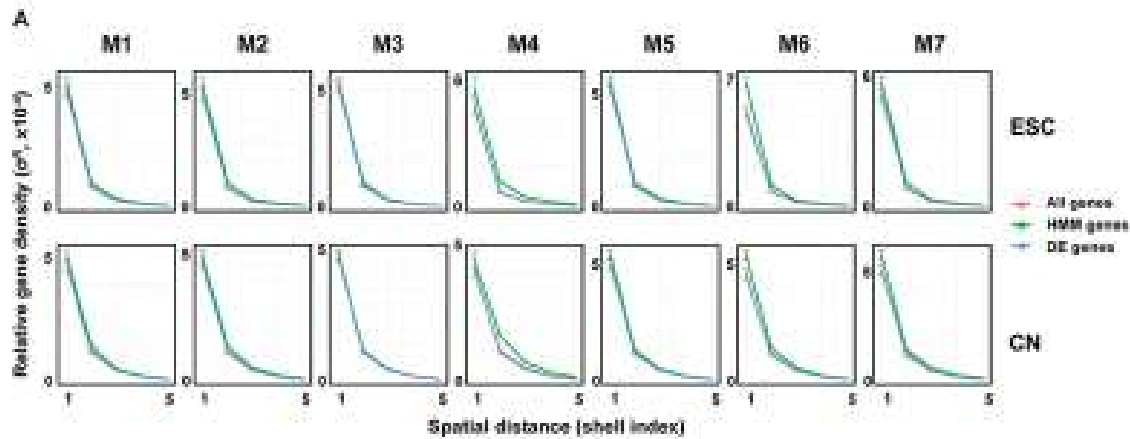

Fig. S4